\documentclass[nofootinbib,aps,a4paper,letterpaper,superscriptaddress,
twocolumn,times,eqsecnum]{revtex4}
\usepackage{amsmath}
\usepackage{amsfonts}
\usepackage{booktabs}
\usepackage{multirow}
\usepackage{siunitx} 
\usepackage{adjustbox}

\usepackage[utf8]{inputenc} 

\usepackage[T1]{fontenc}

\usepackage[greek,english]{babel}

\usepackage{graphicx}
\usepackage{color}
\usepackage{braket}
\usepackage{dcolumn}
\usepackage{bm,url}
\usepackage[linktocpage]{hyperref}
\usepackage{subfigure}
\usepackage{amsfonts}
\usepackage[usenames,dvipsnames,svgnames]{xcolor}  
\usepackage{hyperref}   
\definecolor{oxfordblue}{rgb}{0.0, 0.13, 0.28}
\definecolor{burgundy}{rgb}{0.5, 0.0, 0.13}
\definecolor{darkolivegreen}{rgb}{0.33, 0.42, 0.18}
\definecolor{darkblue}{rgb}{0,0,0.5}
\definecolor{richcarmine}{rgb}{0.84, 0.0, 0.25}
\definecolor{darkblue}{rgb}{0,0,0.5}
\definecolor{bluer}{rgb}{0.00,0.50,0.75}{}
\hypersetup{colorlinks=true, citecolor=red, linkcolor=blue,
 urlcolor = magenta, filecolor=magenta}

\begin{document}
 
 \newcommand\be{\begin{equation}}
  \newcommand\ee{\end{equation}}
 \newcommand\bea{\begin{eqnarray}}
  \newcommand\eea{\end{eqnarray}}
 \newcommand\bseq{\begin{subequations}} 
  \newcommand\eseq{\end{subequations}}
 \newcommand\bcas{\begin{cases}}
  \newcommand\ecas{\end{cases}}
 \newcommand{\p}{\partial}
 \newcommand{\f}{\frac}

 \title{From Cosmology to Cosmonomy}

 \author{Emmanuel N. Saridakis}
 \email{msaridak@noa.gr}
 \affiliation{Institute for Astronomy, Astrophysics, Space Applications and 
Remote Sensing, National Observatory of Athens, 15236 Penteli, Greece}

 \begin{abstract}
For most of its history, cosmology was a qualitatively constrained discourse on 
the universe, shaped by limited observational access and the absence of global 
dynamical laws. This situation has changed decisively in recent decades. Modern 
cosmology is now driven by an unprecedented flow of high-precision data from a 
wide range of independent probes, including the cosmic microwave background, 
large-scale structure, supernovae, baryon acoustic oscillations, gravitational 
lensing, cosmic chronometers, redshift-space distortions, gravitational-wave 
standard sirens, and emerging 21-cm observations, among others.
This observational wealth is matched by a concrete theoretical and 
mathematical framework, 
based on general relativity, which provides the dynamical equations governing 
the evolution of spacetime and matter at cosmic scales. Combined with explicit 
background and perturbative equations, this framework enables quantitative, 
predictive, and falsifiable descriptions of cosmic evolution. Thus, cosmology 
operates today as a nomological natural science of the observable 
universe, characterized by general laws, predictive power, and systematic 
empirical testing. We 
argue that this epistemic 
transformation motivates a corresponding conceptual shift, directly analogous 
to the historical transition from astrology to astronomy. In this sense, the 
transition from cosmology to \emph{cosmonomy} should begin to be discussed 
among cosmologists, or, more precisely, among cosmonomers.

\end{abstract}

 \maketitle
\section{Introduction}

For most of its history, cosmology occupied a peculiar position among the 
natural sciences. 
Although it addressed the universe as a whole, its methodological status 
remained fundamentally distinct from that of disciplines grounded in 
experimentally repeatable phenomena or in strictly testable laws. 
This historical condition is already encoded in terminology. The term 
\emph{cosmology} did not exist in ancient Greece, but was artificially 
introduced in the seventeenth century from the Greek words \emph{kosmos} and 
\emph{logos}, denoting a rational discourse or account of the cosmos rather than 
a science founded on precise and universal laws. 
Indeed, until relatively recently, cosmological research was shaped primarily by 
qualitative reasoning, philosophical consistency arguments, and theoretical 
plausibility, with observational input playing a secondary and often 
non-decisive role.

This situation should not be interpreted as a deficiency, but rather as a 
faithful reflection of the epistemic limitations intrinsic to the subject 
itself. 
The universe is a unique system, inaccessible to controlled experimentation, and 
for a long time it resisted any formulation in terms of strict, predictive, and 
falsifiable laws. 
As a consequence, cosmological inquiry remained naturally intertwined with 
philosophy, metaphysics, and foundational reflection, a status that persisted 
well into the twentieth century. 
Even after the advent of general relativity, which provided the first consistent 
dynamical framework for describing the universe as a whole, cosmology continued 
to be characterized by a plurality of viable models with limited empirical 
discrimination \cite{Weinberg1972,Peebles1993}.

Over roughly the last three decades, however, this situation has undergone a 
profound transformation. 
The emergence of high-precision cosmological observations has altered not only 
the methodology of the field, but also its epistemic status. 
Measurements of the cosmic microwave background anisotropies 
\cite{Smoot1992,Bennett2003,Planck2018}, large-scale structure surveys 
\cite{Tegmark2004}, type Ia supernova observations 
\cite{Riess1998,Perlmutter1999}, baryon acoustic oscillations 
\cite{Eisenstein2005,Alam2017}, weak and strong gravitational lensing 
\cite{Kilbinger2015,DES2018}, and, more recently, gravitational-wave standard 
sirens \cite{Abbott2017}, have collectively established cosmology as a 
quantitatively constrained, predictive, and internally cross-checked scientific 
enterprise. 
Today, cosmological parameters are routinely inferred with percent-level or 
sub-percent-level precision, and theoretical models are systematically 
confronted with mutually independent datasets.

As a result, contemporary cosmology now operates in a regime that differs 
fundamentally from its historical predecessor. 
Its core theoretical structures no longer function merely as frameworks for 
organizing qualitative narratives about the universe, but as effective, law-like 
descriptions whose validity can be tested, constrained, and potentially 
falsified. 
The Friedmann equations, the perturbation theory, and the statistical 
description of 
cosmic structures, now act as operational laws governing the large-scale 
behavior of spacetime and matter, in close analogy with the role played by 
dynamical laws in other mature branches of physics. 
From this perspective, current tensions in cosmological data, such as those 
involving the Hubble parameter or the amplitude of matter fluctuations, should 
not be viewed as signs of conceptual weakness, but rather as manifestations of a 
field that has attained sufficient precision to expose its own limitations and 
falsifiability \cite{CosmoVerseNetwork:2025alb}.

In this work, we argue that this epistemic transformation motivates a 
corresponding conceptual and terminological reassessment. 
In close analogy with the historical transition from \emph{astrology} to 
\emph{astronomy}, which marked the passage from qualitative interpretation to 
law-based description of celestial phenomena, contemporary cosmology has 
effectively crossed the threshold from a \emph{logos} of the cosmos, namely a 
rational 
and discursive account, to a \emph{nomos} of the cosmos, understood as a 
law-governed description. 
We therefore suggest that the term \emph{cosmonomy} may more accurately capture 
the present and emerging status of the field, emphasizing its law-based, 
predictive, and falsifiable character. 
This proposal is not intended as a revision of historical terminology, nor as a 
dismissal of the philosophical depth of cosmological inquiry, but rather as a 
conceptual clarification reflecting the maturation of the discipline.

Accordingly, the aim of this paper is not to introduce new cosmological models 
or observational analyses, but to examine, from a historical, epistemological, 
and methodological perspective, whether contemporary cosmological practice has 
reached a stage at which it can be meaningfully regarded as a law-based science 
of the observable universe as a whole. 
In this context, we propose the term \emph{cosmonomy} as a natural descriptor of 
this stage, together with the corresponding term \emph{cosmonomer} for the 
practitioner engaged in this endeavor. 
Whether or not this terminology is ultimately adopted by the community, we 
believe that reflecting on the conceptual evolution it encapsulates is both 
timely and instructive.

\section{The Cosmos Before Cosmology: Philosophical Foundations}
\label{sec:cosmos_before_cosmology}

In ancient Greek thought, the term \emph{kosmos} did not primarily designate a 
physical system governed by dynamical laws. 
Rather, it referred to an ordered totality: a structured whole characterized by 
harmony, proportion, and intelligibility. 
Its semantic field encompassed notions of order, arrangement, and ornament 
(the words cosmos and cosmetics have the same root), 
emphasizing coherence and completeness. 
In this sense, \emph{kosmos} expressed the conviction that the totality of what 
exists is not chaotic, but ordered in a manner accessible to rational 
contemplation.

This conception is particularly evident in Plato’s \emph{Timaeus}, where the 
cosmos is presented as a unified and living whole, ordered according to 
intelligible principles and fashioned so as to reflect mathematical harmony 
\cite{PlatoTimaeus}. 
Here, the emphasis is not placed on dynamical evolution in time, but on 
ontological completeness and rational design. 
The cosmos is described primarily as that which \emph{is}, rather than as that 
which \emph{evolves} according to explicit laws of motion. 
Temporal change is acknowledged, yet it remains subordinate to a deeper 
metaphysical order that defines the cosmos as a finished and intelligible 
totality.

A closely related perspective is found in Aristotle’s \emph{On the Heavens}, 
where the universe is treated as a self-contained whole endowed with a 
privileged structure and natural places \cite{AristotleDeCaelo}. 
Although Aristotle introduces concepts of motion and causality, these are 
embedded within a teleological and qualitative framework rather than a 
quantitative, law-based one. 
Celestial motions are regular and eternal, but they are not derived from 
universal dynamical equations applicable across scales. 
Instead, the cosmos is divided into qualitatively distinct regions, each 
governed by its own principles, reflecting an ontological hierarchy rather than 
a unified dynamical system.

In both Plato and Aristotle, the cosmos is therefore not conceived as an object 
amenable to systematic experimentation or to mathematical law in the modern 
sense. 
Rather, it is a whole whose intelligibility arises from its ordered structure 
and internal coherence. 
Understanding the cosmos was thus primarily a philosophical task: to articulate 
its meaning, structure, and place within a broader ontological framework, rather 
than to formulate predictive laws governing its evolution.

From this perspective, the absence of the term \textgreek{κοσμολογία} 
(\emph{kosmologia}) in ancient Greek texts is neither accidental nor merely 
linguistic. 
It reflects a deeper epistemic reality, namely the absence of the conditions 
required for a law-based science of the universe as a whole. 
Although ancient thinkers developed sophisticated and often highly systematic 
accounts of the cosmos, these accounts did not, and could not, constitute a 
\emph{cosmology} in the modern sense of a testable, predictive theory of cosmic 
evolution.

Instead, ancient discourse made use of the words \emph{cosmogony} and 
\emph{cosmography}, around which its treatment of the cosmos was largely 
organized. 
Cosmogony addressed the origin of the cosmos, frequently in mythological or 
metaphysical terms, but occasionally through rational speculation, focusing on 
principles of generation rather than on dynamical laws. 
Cosmography, by contrast, concerned the descriptive ordering of the world, 
mapping its structure without seeking universal equations governing its 
behavior. 
Both approaches were inherently qualitative and narrative, reflecting the 
epistemic tools available at the time.

It is nevertheless important to note that certain early Greek philosophical 
schools already entertained the idea that the cosmos unfolds according to 
necessity or rational order. 
Heraclitus, in particular, conceived the world as structured by a universal 
\emph{logos}, while the atomists, most notably Leucippus and 
Democritus, maintained that nothing occurs at random, but that everything 
follows 
from necessity. 
These conceptions, however, remained metaphysical commitments rather than 
empirically grounded or quantitatively formulated laws, and therefore did not 
yet amount to a nomological science of the cosmos.

Crucially, the universe was understood as a unique and singular entity, lacking 
the repeatability and statistical accessibility required for empirical law 
formulation. 
Without the possibility of controlled variation, ensemble reasoning, or 
systematic observation across comparable systems, the formulation of universal 
laws governing the cosmos as a whole remained epistemically inaccessible. 
Even where regularities were identified, they were interpreted as expressions of 
metaphysical necessity or divine order, rather than as empirical laws subject to 
falsification.

From this point of view, the historical absence of cosmology should not be seen 
as a failure of ancient thought, but as a faithful reflection of the epistemic 
status of the subject itself. 
The cosmos, considered as a totality, lay beyond the reach of law-based 
treatment. 
Its uniqueness precluded experimental manipulation, while its immense spatial 
and temporal scales severely limited observational access. 
As a result, discourse about the cosmos necessarily remained at the level of 
rational reflection, philosophical interpretation, and qualitative synthesis. 
The cosmos could be meaningfully contemplated, but not scientifically 
legislated.

\section{From Astrology to Astronomy}
\label{sec:astrology_to_astronomy}

The epistemic situation discussed above for the cosmos as a whole stands in 
sharp contrast with the study of individual celestial bodies within the cosmos. 
The regularity of their motions made them accessible to systematic observation 
and, eventually, to mathematical treatment. 
It is in this restricted but crucial domain that astronomy first emerged as a 
precision science. 
In this section, we examine how the transition from qualitative interpretation 
to law-based description unfolded in the study of the heavens.

\subsection{Celestial regularity and the discovery of laws}
\label{subsec:celestial_regularity}

The decisive factor that enabled the study of the heavens to evolve into a 
law-based science was the exceptional degree of regularity exhibited by 
celestial phenomena. 
Unlike most terrestrial processes, the motions of celestial bodies display 
pronounced periodicity, long-term stability, and repeatability over timescales 
vastly exceeding those of everyday experience. 
These features rendered celestial phenomena uniquely amenable to systematic 
observation, mathematical description, and, ultimately, reliable prediction.

Already in early antiquity, the recurrence of eclipses, conjunctions, and 
planetary cycles made it clear that celestial motions obey stable and persistent 
patterns. 
Even in the absence of explicit physical mechanisms, such regularities could be 
exploited to predict future events with remarkable accuracy. 
This predictive success marked a crucial epistemic transition: celestial 
phenomena were no longer merely described or symbolically interpreted, but 
increasingly treated as manifestations of underlying regularities that could be 
encoded mathematically.

It is important to recall that, in early Greek usage, the terms 
\emph{\textgreek{αστρολογία}} (\emph{astrologia}) and 
\emph{\textgreek{αστρονομία}} (\emph{astronomia}) were not sharply 
distinguished. 
Historically, \emph{astrologia} appears as the earlier and more general 
designation for rational discourse on the heavens, whereas \emph{astronomia} 
emerged later as a more specialized term emphasizing the ordering and law-like 
regularity of celestial motions. 
In their original context, astrology denoted a rational account of the heavens, 
while astronomy highlighted their structured and predictable behavior. 
Neither term carried the metaphysical or divinatory/zodiac connotations 
commonly 
associated with astrology in modern usage. 
Rather, both referred, with different emphases, to what would now be recognized 
as a mathematical and physical study of celestial phenomena.

Thus, the gradual differentiation between astrology and astronomy   reflects 
not a 
change in subject matter, but a shift in epistemic criteria. 
During the Hellenistic period, Greek mathematicians and astronomers, including 
Autolycus, Eudoxus, Callippus, Apollonius, Conon, Hipparchus, and 
Sosigenes, systematically formalized astronomy as a mathematical discipline. 
They developed geometrical models, kinematic schemes, and predictive techniques 
aimed at quantitatively accounting for celestial motions. 
Nevertheless, in parallel intellectual currents and under the influence of 
Babylonian astral traditions, alongside this mathematical consolidation, a 
bifurcation emerged between the exact, mathematical treatment of celestial 
motions and conjectural interpretations concerning their influence on 
terrestrial affairs.
This distinction was articulated with particular clarity by Ptolemy, who 
separated the demonstrative mathematical science of the heavens, developed in 
the \emph{Almagest} \cite{PtolemyAlmagest}, from the probabilistic and 
interpretive framework of judicial astrology presented in the \emph{Tetrabiblos} 
\cite{Tetrabiblos}. 
From this point onward, astronomy increasingly came to denote a discipline 
defined by exact prediction and empirical accountability.

At this stage, the emerging ``science of the stars'' remained largely 
descriptive in 
its physical interpretation. 
Geometrical constructions were devised to reproduce observed motions, without 
necessarily claiming to identify their underlying causes. 
Nevertheless, the very possibility of representing celestial motions through 
mathematical relations signaled a profound shift. 
Once the positions of celestial bodies could be expressed as functions of time,
\begin{equation}
\mathbf{r}_i = \mathbf{r}_i(t),
\end{equation}
the heavens ceased to be objects of qualitative contemplation and became systems 
governed by reproducible and testable patterns.

The key epistemic point is that repeatability made falsification possible. 
Predictions concerning future configurations of celestial bodies could be 
confronted with observation, and persistent discrepancies demanded revision of 
the underlying models. 
In this sense, the transition from astrology to astronomy was not primarily a 
semantic or sociological reclassification, but the emergence of a law-based 
description of nature grounded in regularity, predictability, and empirical 
constraint.

\subsection{Astronomy as the first exact natural science}
\label{subsec:astronomy_exact_science}

As discussed above, Ptolemy’s \emph{Almagest} marks a clear stage in the 
transition from astrology to astronomy, in which celestial motions are treated 
as objects of precise mathematical calculation. 
Although the Ptolemaic system remained kinematic rather than dynamical, its 
predictive power was undeniable. 
Planetary positions, eclipses, and cycles could be calculated with sufficient 
accuracy to establish astronomy as a quantitatively reliable discipline. 
Such predictive control rendered celestial motions amenable not only to 
calculation, but also to mechanical realization, as exemplified by devices such 
as the Antikythera Mechanism, which physically embodied astronomical laws 
\cite{Freeth2006}.

Despite the extraordinary achievements of ancient scholars in describing the 
regularities of the heavens, the decisive step toward a fully nomological 
science was taken in the early modern period. 
This transition was enabled by technological advances, most notably the 
invention of the telescope, which made high-accuracy observations possible, 
and by the formulation of Kepler’s laws, later 
unified within the framework of Newtonian mechanics. 
Kepler’s empirical laws reduced the apparent complexity of planetary motion to 
simple mathematical relations, while Newton demonstrated that these relations 
followed from universal dynamical principles. 
In Newtonian gravity, celestial motion is governed by equations of the form
\begin{equation}
m_i \ddot{\mathbf{r}}_i = - \sum_{j \neq i} G \frac{m_i m_j}{|\mathbf{r}_i - 
\mathbf{r}_j|^3} (\mathbf{r}_i - \mathbf{r}_j),
\end{equation}
which apply universally, independently of the specific nature or location of the 
bodies involved.

With this development, astronomy became the first exact natural science in the 
modern sense. 
Celestial phenomena were no longer explained through ad hoc geometrical 
constructions or metaphysical principles, but derived from universal laws 
capable of unifying diverse observations within a single theoretical framework. 
Prediction, explanation, and falsification became inseparable components of the 
same enterprise. 
This transformation provided the epistemic justification for the term 
\emph{astronomia} as a genuine \emph{nomos} of the stars. 
The heavens could now be understood as obeying strict, quantitative, and 
universally valid laws. 
The significance of this achievement was not merely technical, but conceptual: 
it established that at least part of the natural world admits a law-based 
description that is both predictive and empirically testable.

\section{The birth and long maturity of cosmology}
\label{sec:birth_cosmology}

Despite the success of astronomy, the extension of a fully nomological framework 
to the observable universe as a whole remained, for a long time, out of reach. 
The cosmos differs from individual celestial systems in ways that are 
epistemically decisive. 
Most importantly, it is a singular object: there is only one universe, which 
cannot be varied, replicated, or subjected to controlled experimental 
manipulation.

As a consequence, the methodological conditions that support law formulation in 
astronomy (repeatability, ensemble reasoning, and controlled comparison) were 
absent when the cosmos was considered as a totality. 
Even if large-scale regularities were suspected, there was no clear sense 
where and how they could be tested or falsified. 
Without access to multiple realizations or statistically independent samples, 
the universe could not be treated as a system governed by empirically 
established laws.

\subsection{The invention of ``cosmology'' in early modern thought}
\label{subsec:invention_cosmology}

The emergence of the term \emph{cosmology} marks an important conceptual moment 
in the intellectual history of the study of the universe. 
Nevertheless, its appearance does not signal the maturation of an already 
law-based science, but rather the formalization of a metaphysical domain 
concerned with the universe taken as a whole.

The earliest known occurrence of the word \emph{cosmology} dates to the 
seventeenth century. 
In English, it is attested in Thomas Blount’s \emph{Glossographia} (1656), where 
it is defined as a discourse concerning the world or the universe 
\cite{Blount1656}. 
This lexical introduction is revealing: the term enters scientific and 
philosophical language not as the name of an established empirical discipline, 
but as a label for a general rational treatment of the cosmos.

The systematic introduction of \emph{cosmologia} as a distinct branch of 
knowledge is due to Christian Wolff in the early eighteenth century. 
In his \emph{Cosmologia Generalis} (1731), Wolff explicitly classifies cosmology 
as a subdivision of metaphysics, alongside ontology, rational psychology, and 
natural theology \cite{Wolff1731}. 
Within this framework, cosmology concerns the most general properties of the 
world as a whole, abstracted from particular physical processes and independent 
of empirical measurement.

This classification is epistemically significant. 
Cosmology is introduced neither as an extension of astronomy nor as a natural 
science grounded in observation, but as a rational inquiry into the world 
considered in its totality. 
Its subject matter is defined by maximal generality rather than by dynamical 
specificity. 
Accordingly, early cosmology inherits the structural features of philosophical 
discourse: internal coherence, conceptual clarity, and logical consistency take 
precedence over empirical confrontation.

At this stage, cosmology is explicitly non-empirical. 
The universe is treated as a necessary object of reason, not as a system 
governed by measurable laws. 
This status is not accidental, but reflects the absence of observational access 
to global cosmic properties. 
Without data capable of constraining cosmic structure or evolution, cosmology 
could not yet aspire to the status of a law-based science. 
For these reasons, discourse about the cosmos remained outside the domain of 
\emph{nomos} and belonged instead to \emph{logos}: rational reflection on the 
structure and meaning of the totality of existence.

\subsection{Cosmology before the precision era}
\label{subsec:cosmology_pre_precision}

The transition from metaphysical cosmology to a proto-scientific discipline 
begins only in the twentieth century, with the advent of relativistic gravity. 
General relativity provided, for the first time, a consistent theoretical 
framework within which the observable universe could be treated as a dynamical 
entity 
\cite{Hetherington2014,Saridakis:2021qzv}. 
Cosmological models developed by Friedmann, Lemaître, and later Robertson and 
Walker opened the possibility of describing the large-scale structure and 
evolution of spacetime in explicit mathematical terms.

Nevertheless, for several decades cosmology remained only weakly constrained by 
observation. 
Although theoretical models proliferated, empirical discrimination between them 
was limited. 
Early observations, including galaxy redshift measurements and estimates of 
cosmic expansion, offered suggestive indications rather than decisive tests. 
As a result, a wide variety of cosmological scenarios coexisted, often differing 
in fundamental assumptions concerning geometry, matter content, and cosmic 
history.

This period is marked by a striking plurality of models. 
Different cosmological solutions of Einstein’s equations were explored primarily 
on theoretical grounds, with observational data playing a secondary and often 
ambiguous role. 
In the absence of precise measurements, cosmological reasoning relied heavily on 
simplicity arguments, philosophical preferences, and internal theoretical 
consistency. 
The universe could be modeled, but not yet legislated.

Consequently, cosmology during this era occupied an intermediate epistemic 
position. 
It was no longer purely metaphysical, yet it had not acquired the methodological 
rigor characteristic of a mature natural science. 
Its status was that of a boundary discipline, situated between physics and 
philosophy, between mathematical formalism and conceptual speculation. 
This liminal character is reflected in persistent debates concerning initial 
conditions, global geometry, and the interpretation of cosmological solutions. 
Accordingly, cosmology in this period was typically presented as a late and 
conceptually oriented application of general relativity, often appearing as a 
concluding chapter in general relativity textbooks rather than as an autonomous 
empirical science.

\section{The data revolution and the emergence of cosmic laws}
\label{sec:data_revolution}

With the advent of high-precision observational capabilities, cosmology 
gradually began to shed the intermediate epistemic status that had long 
characterized the field. 
Advances in astronomical instrumentation, survey design, and data analysis 
opened observational access to the universe across a wide range of physical 
scales and cosmic epochs, transforming what had previously been a weakly 
constrained theoretical enterprise into an empirically grounded science. 
This extended process of maturation ultimately led to a qualitative shift: 
cosmology moved from a largely speculative or interpretive activity toward a 
quantitatively testable, law-based, and data-driven description of the 
universe. 
It is this transformation, driven by the convergence of precise observations 
and 
well-defined theoretical frameworks, that forms the subject of the present 
section and motivates the conceptual shift discussed thereafter.

\subsection{Cosmic dynamics as a law-based description of the Universe}
\label{subsec:cosmic_dynamics}

The decisive step that elevates modern cosmology from a descriptive framework to 
a genuinely law-based science is the existence of explicit dynamical equations 
governing the universe as a whole. 
Within general relativity, the large-scale dynamics of spacetime and matter are 
encoded in a compact set of equations whose domain of validity extends across 
the entire observable universe.

Starting from the observationally motivated assumption that the Universe is 
homogeneous and isotropic on sufficiently large scales (the Cosmological 
Principle), general relativity admits a quantitative description based on the 
Friedmann-Lemaître-Robertson-Walker spacetime metric 
\begin{equation}
ds^{2} = -dt^{2} + a^{2}(t)\left[\frac{dr^{2}}{1 - k r^{2}} + 
r^{2}\left(d\theta^{2} + \sin^{2}\theta\, d\phi^{2}\right)\right],
\end{equation}
where $a(t)$ is the scale factor and $k = +1,0,-1$ denotes closed, flat, or open 
spatial geometry. 
Within this framework, cosmic evolution is governed by the Friedmann equations 
\begin{equation}
H^2 \equiv \left(\frac{\dot a}{a}\right)^2 = \frac{8\pi G}{3}\rho - 
\frac{k}{a^2} + \frac{\Lambda}{3},
\end{equation}
\begin{equation}
\frac{\ddot a}{a} = -\frac{4\pi G}{3}(\rho + 3p) + \frac{\Lambda}{3},
\end{equation}
supplemented by the continuity equation,
\begin{equation}
\dot{\rho} + 3H(\rho + p) = 0,
\end{equation}
where $H=\dot{a}/a$ is the Hubble function, $\rho$ and $p$ denote the energy 
density and pressure of the cosmic fluid, and $\Lambda$ is the cosmological 
constant. 
Together, these relations describe the global expansion of the universe and 
enable quantitative predictions for its past and future evolution once the 
properties of the cosmic components are specified.

Beyond the background level, cosmology admits a systematic perturbative 
expansion that governs the formation and evolution of cosmic structures. 
For scalar perturbations in the linear regime, the evolution of matter 
overdensity $
\delta \equiv \frac{\delta \rho}{\rho}$
is described, in the sub-horizon limit, by equations of the form
\begin{equation}
\ddot{\delta} + 2H\dot{\delta} - 4\pi G \rho\,\delta = 0,
\end{equation}
while the full relativistic treatment involves coupled equations for metric and 
matter perturbations.
The solutions of these equations predict the growth of structure, gravitational 
potentials, and anisotropies across cosmic time.

These perturbative relations constitute the theoretical backbone of large-scale 
structure formation. 
They allow cosmology to produce quantitative predictions for galaxy clustering, 
weak gravitational lensing, redshift-space distortions, and the anisotropies of 
the cosmic microwave background. 
Crucially, these predictions are not qualitative trends, but precise relations 
among observables that can be directly confronted with data.

\subsection{Observational probes and datasets of precision cosmology}
\label{subsec:observational_probes}

The nomological character of modern cosmology would remain purely formal in the 
absence of observational access to the quantities appearing in the cosmic 
dynamical equations. 
Over the last decades, a remarkably rich and diverse set of independent yet 
complementary observational probes has emerged, enabling precision tests of 
cosmic laws across a wide range of redshifts and physical scales.

\paragraph{Cosmic Microwave Background (CMB).}
Measurements of temperature and polarization anisotropies of the cosmic 
microwave background, from COBE to WMAP and Planck, provide a snapshot of the 
universe at recombination and encode information about primordial perturbations, 
geometry, and cosmic composition \cite{Smoot1992,Bennett2003,Planck2018}. 
The CMB remains one of the most precise datasets in all of physics.

\paragraph{Type Ia Supernovae (SNIa).}
Observations of type Ia supernovae serve as standardized candles, enabling 
direct reconstruction of the expansion history of the universe and providing the 
first robust evidence for late-time cosmic acceleration 
\cite{Riess1998,Perlmutter1999}.

\paragraph{Baryon Acoustic Oscillations (BAO).}
BAO measurements in galaxy surveys define a standard ruler that constrains the 
expansion rate and angular diameter distance as functions of redshift, offering 
a geometrical probe complementary to supernova observations 
\cite{Eisenstein2005,Alam2017}.

\paragraph{Large-Scale Structure and Galaxy Clustering.}
Large redshift surveys such as 2dF, SDSS, BOSS/eBOSS, and DES measure the 
statistical distribution of galaxies, constraining the matter density, galaxy 
bias, and the growth of cosmic structure \cite{Tegmark2004}.

\paragraph{Cosmic Chronometers.}
Differential age measurements of passively evolving galaxies provide direct 
determinations of the Hubble function $H(z)$, independent of integrated 
distance indicators \cite{Jimenez2002,Moresco2016}.

\paragraph{Redshift-Space Distortions and $f\sigma_8$ Measurements.}
Velocity-induced anisotropies in galaxy clustering yield direct access to the 
growth rate of structure through measurements of $f\sigma_8$, enabling tests of 
gravity on cosmological scales \cite{Song2009}.

\paragraph{Weak and Strong Gravitational Lensing.}
Gravitational lensing probes the integrated matter distribution and spacetime 
geometry, offering powerful tests of structure formation and gravitational 
dynamics \cite{Kilbinger2015,DES2018}.

\paragraph{Galaxy Clusters.}
Cluster number counts, mass functions, and internal properties constrain the 
amplitude of matter fluctuations and the growth of structure, linking cosmic 
expansion to nonlinear gravitational collapse \cite{Allen2011}.

\paragraph{Gravitational Waves (Standard Sirens).}
Compact binary mergers detected through gravitational waves provide absolute 
luminosity distances, offering an independent probe of cosmic expansion that 
does not rely on the cosmic distance ladder \cite{Abbott2017}.

\paragraph{21-cm Cosmology.}
Observations of the hyperfine transition of neutral hydrogen promise to probe 
the dark ages, cosmic dawn, and reionization, potentially extending precision 
cosmology to very high redshifts \cite{Furlanetto2006}.

Taken together, these probes form an interconnected observational network. 
They do not merely measure isolated quantities, but collectively test the 
internal consistency of the cosmic dynamical framework across epochs, length 
scales, and physical regimes. 
Moreover, this network does not operate as a one-directional verification 
mechanism. 
Instead, it establishes a dialectical interplay: precision observations reveal 
tensions and inconsistencies, such as the Hubble tension, that motivate the 
development of new theoretical syntheses, which in turn generate novel 
observational predictions and drive further experimental and methodological 
advances.

\subsection{Cosmology as a falsifiable and self-consistent law-based science}
\label{subsec:falsifiable_cosmology}

The defining feature of contemporary cosmology is not the mere existence of 
equations, but their systematic confrontation with an extensive and 
heterogeneous body of data. 
Cosmological parameters are now inferred through global likelihood analyses that 
combine multiple datasets, yielding constraints of unprecedented precision.

Equally important is the internal cross-consistency of cosmological inference. 
Independent probes constrain overlapping combinations of parameters, enabling 
nontrivial consistency tests of the underlying theoretical framework. 
Discrepancies, such as those involving the Hubble constant or the amplitude of 
matter fluctuations \cite{CosmoVerseNetwork:2025alb}, arise precisely because 
cosmology has reached the level of precision required to expose the limitations 
of its effective laws.

In this sense, modern cosmology satisfies the operational criteria of a 
nomological science, including the formulation of general laws, predictive 
power, and systematic empirical testing 
\cite{Hempel1965,Salmon1984,vanFraassen1980,Cartwright1999,Norton2003}. 
It possesses universal equations, predictive structures, and a dense network of 
observational tests. 
The observable universe, once accessible primarily through philosophical 
reflection, has thus become an object of quantitative legislation.

It is this empirical, mathematical, and epistemic transformation that motivates 
the proposal to regard contemporary cosmology not merely as a \emph{logos} of 
the cosmos, but as a genuine \emph{nomos} of the cosmos.

\section{From Cosmology to Cosmonomy}
\label{sec:cosmonomy}

The developments discussed in the previous sections invite a reassessment of 
the conceptual status of contemporary cosmology. 
Historically, cosmology was appropriately understood as a \emph{logos} of the 
cosmos: a rational discourse aimed at organizing, interpreting, and 
philosophically situating the universe as a whole. 
This designation   reflected both the nature of its subject matter and 
the epistemic tools available at the time. 
In the absence of global dynamical equations and precision observations, 
cosmology could not reasonably aspire to the formulation of laws in the strict 
sense.

This situation has now changed in a fundamental way. 
As shown in Section~\ref{sec:data_revolution}, modern cosmology rests on a 
well-defined set of dynamical equations governing cosmic evolution, supplemented 
by a perturbative framework that predicts the formation and growth of 
structure. 
These equations are not merely formal constructions. 
They are systematically confronted with a broad and diverse body of 
observational data, and their parameters are inferred with increasingly high 
precision. 
In operational terms, they function as effective laws of nature.

At a deeper conceptual level, this transition mirrors the historical evolution 
from astrology to astronomy. 
Astronomy acquired its status as a ``\emph{nomy}'' when celestial phenomena 
were 
shown to obey universal, predictive, and empirically testable laws. 
Once planetary motions could be derived from dynamical principles and subjected 
to falsification, the study of the heavens ceased to rely on qualitative 
interpretation and became an exact natural science. 
In an analogous manner, the observable universe itself has now become an object 
of quantitative legislation.

From an etymological perspective, this shift can be expressed as a passage from 
\emph{logos} to \emph{nomos}. 
The proposal to speak of \emph{cosmonomy} is therefore not a semantic novelty, 
but a conceptual clarification. 
It reflects the fact that the cosmos is now investigated through laws that 
constrain its behavior and evolution, rather than through qualitative narratives 
alone.

In contemporary practice, the study of the observable universe satisfies the 
defining criteria of a nomological science. 
Cosmic evolution is described by equations that are universal in scope, 
predictive in content, and falsifiable in principle. 
These equations relate observables across cosmic time and length scales, 
enabling quantitative reconstruction of the universe’s history and constrained 
extrapolation toward its future.

Moreover, modern cosmology does not rely on a single observational handle. 
Instead, it operates through a tightly interconnected network of probes that 
test the same underlying laws from distinct physical perspectives. 
The consistency between background expansion, structure growth, gravitational 
lensing, and primordial anisotropies is not imposed by assumption, but 
established empirically. 
Where tensions arise, they serve as diagnostics of either unresolved 
observational systematics or the limitations of the effective laws themselves.

In this sense, \emph{cosmonomy} denotes neither a claim of finality nor the 
discovery of immutable truths about the universe. 
Rather, it designates a mode of inquiry in which cosmic laws are treated as 
effective, testable, and revisable descriptions, in close analogy with the laws 
governing other complex physical systems. 
The universe remains unique, but it is no longer epistemically exceptional.

Importantly, cosmonomy is deeply connected to fundamental physics. 
Questions concerning the nature of dark energy, dark matter, gravity, and the 
``initial'' conditions of the universe are now addressed within a framework 
that 
directly links theoretical principles to observation. 
Cosmic laws thus provide a bridge between fundamental physics and empirical 
reality, situating cosmonomy firmly within the core of modern physics.

The nomological character of cosmology is expected to be further strengthened by 
forthcoming observational programs. 
Next-generation surveys and multi-messenger observations will extend precision 
tests of cosmic laws to new regimes, higher redshifts, and previously 
inaccessible physical processes. 
The scope of cosmic legislation will therefore continue to expand, both in depth 
and in breadth.

At the same time, it is essential to recognize the intrinsic limits of law-based 
descriptions. 
Cosmic laws are necessarily effective, reflecting both the finite precision of 
observations and the theoretical assumptions underlying their formulation. 
The existence of horizons, the uniqueness of the universe, and the role of 
initial conditions impose fundamental constraints on what can be known and 
predicted. 
For this reason, a clear distinction must be drawn between the observable 
universe and the totality of reality. 
In this sense, the study of the cosmos should be regarded as a local science, 
while rational accounts of regions beyond the cosmological horizon remain within 
the domain of \emph{logos}. 
Acknowledging these limits is a hallmark of a mature nomological perspective, 
not a refutation of it.

Against this background, the proposal to adopt the term \emph{cosmonomy} should 
be understood as a reflection of the evolving identity of the field. 
It neither denies the historical roots of cosmology nor diminishes the 
philosophical depth of questions concerning the universe as a whole. 
Rather, it highlights the fact that cosmology has entered a stage in which its 
central activity consists in the formulation, testing, and refinement of cosmic 
laws.

\section{Conclusions}
\label{sec:conclusions}

In this work, we have examined the conceptual evolution of cosmology from a 
historical, epistemological, and methodological perspective, focusing in 
particular on the criteria that distinguish a qualitative discourse from a 
law-based scientific discipline. 
Our analysis shows that, for most of its history, cosmology was necessarily 
confined to the level of \emph{logos}: a rational and often profound reflection 
on the universe as a whole, shaped by the absence of global dynamical laws and 
by severe observational limitations. 
This status was neither accidental nor provisional, but a faithful expression of 
the epistemic conditions under which the cosmos could be studied.

Over the last few decades, this situation has changed in a decisive way. 
Contemporary cosmology is now characterized by an unprecedented abundance of 
high-quality data, originating from a wide spectrum of independent and 
complementary observational probes. 
Measurements of the cosmic microwave background, large-scale structure, 
supernovae, baryon acoustic oscillations, cosmic chronometers, gravitational 
lensing, galaxy clusters, redshift-space distortions, gravitational waves, and 
emerging 21-cm observations collectively provide access to the universe across 
cosmic time and physical scale. 
Crucially, these datasets are not merely accumulated, but systematically 
integrated through mature statistical and computational frameworks that enable 
precision parameter estimation, consistency testing, and meaningful model 
discrimination.

At the theoretical level, this observational transformation is matched by a 
well-defined dynamical framework. 
The evolution of the universe is described by explicit equations at both the 
background and perturbative levels, allowing for quantitative predictions of 
expansion history, structure formation, and observable correlations. 
These relations function operationally as effective laws: they are universal in 
scope, predictive in content, and continuously confronted with data. 
The tensions and discrepancies that arise within this framework should therefore 
not be interpreted as signs of conceptual fragility, but as indicators of a 
discipline that has reached the level of precision required to probe the limits 
of its own laws.

Taken together, these developments motivate a reassessment of the conceptual 
status of the field. 
Modern cosmology no longer consists primarily in the qualitative interpretation 
of the cosmos, but in the formulation, testing, and refinement of quantitative 
laws governing its evolution. 
In this sense, the observable universe has ceased to be epistemically 
exceptional. 
Despite its uniqueness, it is now investigated through methods that are 
operationally indistinguishable from those employed in other law-based branches 
of physics.

On this basis, we have suggested that the term \emph{cosmonomy} may provide a 
more accurate descriptor of the present and emerging practice of the field, 
emphasizing its nomological, predictive, and falsifiable character. 
This proposal is not intended to erase the historical meaning of cosmology, nor 
to diminish the philosophical depth of questions concerning the universe as a 
whole, but it reflects the maturation of the discipline itself, and the fact 
that 
cosmology has acquired both the empirical foundation and the theoretical 
machinery required for genuine cosmic legislation.

Whether or not this terminology is ultimately adopted by the community, the 
transformation it seeks to articulate is already under way. 
The observable universe is no longer studied solely as an object of 
contemplation, but as a system governed by empirically grounded and 
quantitatively precise laws, persistently confronted with an ever-growing body 
of data. 
In this precise and limited sense, cosmology has effectively become cosmonomy.

 \begin{acknowledgments}
 The author  acknowledges the 
contribution of the LISA CosWG   and of COST Actions 
CA21136 ``Addressing observational tensions in cosmology with systematics and 
fundamental physics (CosmoVerse)'',  CA21106 ``COSMIC WISPers in the Dark 
Universe: Theory, astrophysics and experiments (CosmicWISPers)'', and CA23130 
``Bridging high and low energies in search of quantum gravity (BridgeQG)''.
 \end{acknowledgments}
 
 
\end{document}